# Extreme-ultraviolet pump-probe studies of one femtosecond scale electron dynamics

P. Tzallas[1*], E. Skantzakis[1,2], L. A. A. Nikolopoulos[3], G. D. Tsakiris[4] and D. Charalambidis[1,2]

[1]*Foundation for Research and Technology—Hellas, Institute of Electronic Structure and Laser, PO Box 1527, GR-711 10 Heraklion, Crete, Greece*

[2]*Department of Physics, University of Crete, PO Box 2208, GR71003 Heraklion, Crete, Greece*

[3]*School of Physical Science, Dublin City University, Glasnevin, Dublin 9, Ireland*

[4]*Max-Planck-Institut für Quantenoptik, Hans-Kopfermann-Str. 1, D-85748 Garching, Germany*

*Corresponding author e-mail address: ptzallas@iesl.forth.gr

**Studies of ultrafast dynamics along with femtosecond-pulse metrology rely on non-linear processes, induced solely by the exciting/probing pulses or the pulses to be characterized. Extension of these approaches to the extreme-ultraviolet (XUV) spectral region opens up a new, direct route to attosecond scale dynamics. Limitations in available intensities of coherent XUV continua kept this prospect barren. The present work overcomes this barrier. Reaching condition at which simultaneous ejection of two bound electrons by two-XUV-photon absorption becomes more efficient than their one-by-one removal it is succeeded to probe atomic coherences, evolving at the 1fs scale, and determine the XUV-pulse duration. The investigated rich and dense in structure autoionizing manifold ascertains applicability of the approach to complex systems. This initiates the era of XUV-pump-XUV-probe experiments with attosecond resolution.**

A large variety of ultrafast phenomena including electronic motion in atoms, molecules, condense matter and plasmas, dynamic electron-electron correlations, charge migration, ultrafast dissociation and reaction processes occur on the few-fs to attosecond temporal scale. Attosecond (asec) pulses [1] provide access to these temporal regimes in different states of matter [2-6]. Non-linear (NL) XUV processes, constitute the ideal tool for the study such dynamics. Attosecond pulse trains [7-9] reached intensities sufficient to induce two-XUV-photon processes [10-14]. However, isolated asec pulses, requisite for XUV-pump-XUV-probe experiments, have not so far attained the required parameters for an observable two-XUV-photon process. As a consequence asec pulse metrology and time domain applications have been widely based on infrared (IR)-XUV cross-correlations approaches entail assumptions for the analysis [15].

The present work succeeds for the first time in observing two-XUV-photon processes induced by energetic XUV continua, temporally confined in isolated pulses with duration of the order of 1fs. These processes are in turn exploited in XUV-pumping-XUV-probing ultra-fast evolving atomic coherences, as well as in determining the duration of the XUV bursts. A structured part of the single continuum of the Xenon atom is excited by the first pulse forming an electronic wave packet that undergoes rapid and complex motion before it decays. These dynamics are traced, thanks to the XUV parameters reached, at which a second pulse ejects a second electron before the first one leaves the atom taking with it all the information on the temporal evolution of the system (coherence decay). Unconventionally, the two electrons leave the atom together and thus doubly ionized Xe yield as a function of the delay between the two pulses carries the fingerprint of the wave packet motion and the XUV pulse duration. Since the pulse duration and the decay time of the wave packet are very different, pulse duration and wave packet dynamics are imprinted in different regions of the measured trace.

The intense XUV radiation is generated by frequency up-conversion of many cycle high peak power laser fields interacting with an atomic target. To obtain the XUV continua the technique of the Interferometric Polarization Gating (IPG) [16, 17] is applied. Since the laser used is not carrier-envelope phase (CEP) stabilized, only a fraction of the laser shots, generate pure XUV continuum spectra and thus single pulses [18]. Pulses with CEP far from $\pi/2$ generate modulated spectra [16, 18] resulting in waveforms with a more or less resolved double peak structure. The discussion to follow will be confined to pure continua. Possible modifications of the results arising from the shot to shot CEP variation will be further considered separately.

The experimental set up used is shown in Fig. 1a. 38 fs long Ti:Sapph pulses enter an IPG device (see supplementary information (SI)). The ellipticity modulated output pulses interact with a Xenon gas jet producing the XUV radiation. The radiation transmitted through a Sn filter spans the spectral range 50-80 nm (Fig. 1b). The XUV radiation is focused into a second gas jet (He or Xe) by a bisected spherical mirror acting also as a wave-front splitter. Translation of one mirror half produces two mutually delayed pulses with variable delay.

Two-XUV-photon processes are observed in two different experiments. In the first, two photon ionization of He was verified by the dependence of the ion signal on the XUV intensity (see SI). In the second doubly ionized Xenon has been observed in the measured mass ion spectra (Fig. 1c). For the given XUV spectral width $Xe^{2+}$ can only be produced through multi-XUV-photon absorption (Fig. 2). The most pronounced possible channels are: I) two-photon direct double ionization (TPDDI) in which single XUV photon absorption is in the vicinity of a manifold of doubly excited $5s^2 5p^4[^3P, ^1D, ^1S]mlm'l'$ and inner-shell excited $5s5p^6(^2S_{1/2})np[^1P_1]$ (Auger) autoionizing states (AIS), from where absorption of a second XUV-photon ejects two electrons that share the excess energy , and II) Single-XUV-photon ionization of Xe, followed by a single-XUV-photon ionization of the two fine structure levels of the $Xe^+$ ground state by the high energy (above 21 or 19.7eV for each level) part of the continuum spectrum distribution that exceeds the double ionization threshold (sequential double ionization (SDI)). Additional less probable sequential channels through excited ionic states are discussed in the SI. While for pulse durations $\tau_{XUV}$ of tens of fs the sequential two-XUV-photon ionization is the dominant process, as pulse duration decreases the direct process gains in relative strength. The reason for this relates to the different dependence on the pulse duration of the direct ($\propto \tau_{XUV}$) and sequential ($\propto \tau_{XUV}^2$) ionization process [19, 20]. At the 1 fs pulse duration level the direct process becomes the strongest double ionization channel, verified experimentally in this work.

Using a sequence of two XUV pulses and varying the delay τ between them, we are able to: a) measure their duration by means of a 2nd order intensity volume autocorrelation (2nd order IVAC), from the part of the trace in which the two pulses are overlapping and b) to induce, control and probe a fast evolving coherence in the structured continuum, evaluating the part of the trace where the two pulses are not overlapping. Here the first pulse (P1) induces the coherence, pumping a coherent superposition of the AIS manifold and ionic states. It also, partially doubly ionizes the atom ejecting part of the formed electron wave-packet. The second pulse (P2): 1) excites a replica of the wave-packet, which interferes with the evolved first one. Thus it controls the excitation by means of constructive or destructive interferences occurring at different frequencies. 2) Partially ionizes the wave-packet it excites and probes the evolution of the first excited wave-packet. The path involving excitation by P1 and double ionization by P2 is to our knowledge the first successful experimental implementation of an XUV-pump-XUV-probe sequence.

The measured trace is shown in Fig. 3a. The pronounced maximum around $\tau=0$ is the 2nd order IVAC trace. This is because contributions to this part is predominantly by the TPDDI channel (a) due to of the shortness of the pulse and (b) because only a small portion of the spectrum contributes to the sequential channel. A Gaussian distribution fit (Fig. 3b) results a pulse duration of 1.5 fs. It is well known [16, 18] that due to the shot to shot CEP variation, the spectrum varies from pure to modulated continuum, which in the time domain translates to a variation between a clean single pulse and pulses with a double maxima structure. Thus the measured duration of 1.5 fs is the average of the duration of single pulses, most probably of sub-fs duration and twin maxima pulses with a peak separation half the laser period (1.33 fs). Consequently the measured duration is $1.5^{+0.2}_{-1.1}$ fs (420 asec being the Fourier Transform Limited (FTL) duration). This duration further verifies the dominance of the TPDDI channel. A SDI trace would be the cycle-average of the square of the first order autocorrelation of the field and thus would have a width equal to the width of the FT of the XUV spectrum (420 asec). The much larger measured width confirms that the two electrons are ejected “together” before the system finds the time to first decay in $Xe^+ + e^-$.

The beating signal at delays > 5 fs, results from the XUV-pump-XUV-probe process of the atomic coherence. At the temporal resolution of the experiment high frequency Ramsey fringes corresponding to the excitation frequencies of each AIS are averaged out and the low frequency components of the evolution of the coherent superposition dominate the trace [21]. Contributions from modulations in the excitation process due to the two interfering wave-packets are not to be excluded [22, 23]. Fourier Transform (FT) of the traces reveals frequency differences of the excited states (Fig. 3c). An assignment of the FT spectrum peaks is given in the SI. The most of the peak positions coincide, within the error, with those measured in ref. 5, where the dynamics were observed at the autoionization process. Double peak structure of some XUV pulses, due to the variation of the CEP, modifies the exciting amplitude distribution and introduces in the trace components shifted by half the laser period, which only reduces the fringe contrast in the averaged trace.

In a similar study involving only bound states and IR fs pulses, a theoretical description is given in the weak field approximation [21]. At the present conditions, this approximation might not be valid, as ground state depletion and higher order couplings cannot be excluded. Since calculation of the relevant Xenon atomic structure is not tractable we present ab-initio time-dependent perturbative calculations of the appropriate order in Helium (see SI).

In summary, the present work establishes the era of XUV-pump-XUV-probe experiments at the 1fs temporal scale, along with the NL-XUV process based isolated attosecond pulse metrology. The rich and dense in structure investigated part of the spectrum signifies applicability of the approach to complex systems. In this sense the present proof of principle experiment opens up a new chapter in time domain studies of realistic complex systems, at ultra-high temporal resolution.

**Acknowledgments**

This work is supported in part by the European Commission programmes ULF, ALADIN (GA-228334), ATTOFEL (GA-238362), FASTQUAST (PITN-GA-2008-214962), ELI-PP (GA-212105) and FLUX program (PIAPP-GA-2008-218053) of the

7th FP. L.A.A.N. acknowledges support from COST CM0702 action and ICHEC at Dublin.

## References

1. Krausz, F., Ivanov, M., Attosecond Physics. ,*Rev. Mod. Phys.* **81**, 163 (2009).
2. Remetter T. *et al.*, Attosecond electron wave packet interferometry. *Nature Phys.* **2**, 323 (2006).
3. Cavalieri, A. L. et al., Attosecond spectroscopy in condensed matter. *Nature* **449**, 1029 (2007).
4. Sansone, G. *et al.*, Electron localization following attosecond molecular photoionization. *Nature* **465,** 763 (2010).
5. Skantzakis, E. *et al.*, Tracking Autoionizing-Wave-Packet Dynamics at the 1-fs Temporal scale. *Phys. Rev. Lett.* **105**, 043902 (2010).
6. Goulielmakis, E. *et al.*, Real-time observation of valence electron motion. *Nature* **466**, 739 (2010).
7. Tzallas, P. *et al.*, Direct observation of attosecond light bunching. *Nature* **426**, 267 (2003).
8. Nabekawa, Y. *et al.*, Conclusive evidence of an attosecond pulse train observed with the mode-resolved autocorrelation technique. *Phys. Rev. Lett.* **96**, 083901 (2005).
9. Nomura, Y. *et al.*, Attosecond phase locking of harmonics emitted from laser-produced plasmas. *Nature Phys.* **5**, 124 (2009).
10. Xenakis, D. *et al.*, Observation of two-XUV-photon ionization using harmonic generation from a short, intense laser pulse. *J. Phys.* B **29**, L457 (1996).
11. Papadogiannis, N. A. *et al.*, Two-Photon Ionization of He through a Superposition of Higher Harmonics. *Phys. Rev. Lett.* **90**, 133902 (2003).
12. Sekikawa, T. *et al.*, Pulse Compression of a High-Order Harmonic by Compensating the Atomic Dipole Phase. *Phys. Rev. Lett.* **83**, 2564 (1999).

13. Sekikawa, T. *et al.*, Measurement of the Intensity-Dependent Atomic Dipole Phase of a High Harmonic by Frequency-Resolved Optical Gating. *Phys. Rev. Lett.* **88**, 193902 (2002).

14. Nabekawa, Y. *et al.*, Production of Doubly Charged Helium Ions by Two-Photon Absorption of an Intense Sub-10-fs Soft X-Ray Pulse at 42 eV Photon Energy. *Phys. Rev. Lett*. **94**, 043001 (2005).

15. Kruse, J. *et al.*, Inconsistencies between two attosecond pulse metrology methods: A comparative study. *Phys. Rev.* A **82**, 021402(R) (2010).

16. Tzallas, P. *et al.*, Generation of intense continuum extreme-ultraviolet radiation by many cycle laser fields. *Nature Phys.* **3**, 846 (2007).

17. Skantzakis, E. *et al.*, Coherent continuum extreme ultraviolet radiation in the sub 100-nJ range generated by a high-power many-cycle laser field. *Opt. Lett.* **34**, 1732 (2009).

18. Tzallas, P. *et al.*, Measuring the absolute carrier-envelope phase of many-cycle laser fields. *Phys. Rev. A* **82**, 061401(R) (2010).

19. Lambropoulos, P. *et al.*, Direct versus sequential double ionization in atomic systems. *Phys. Rev. A* **78**, 055402 (2008).

20. Nakajima, T., L. A. A. Nikolopoulos, Use of helium double ionization for autocorrelation of an xuv pulse. *Phys. Rev. A* **66**, 041402(R) (2002).

21. Blanchet, V. *et al.*, Temporal coherent control in the photoionization of Cs2: Theory and experiment. *J. Chem. Phys.* **108**, 4862 (1998).

22. Campbell, M. B. *et al.*, Observation of oscillations between degenerate bound-state configurations in rapidly autoionizing two-electron atoms. *Phys. Rev. A* **57**, 4616 (1998).

23. Jones, R. R. *et al.*, Ramsey interference in strongly driven Rydberg systems. *Phys. Rev. Lett.* **71**, 2575 (1993).

**Figure Captions**

**Figure 1. Schematic diagram showing the XUV-pump-XUV-probe apparatus** (**a**) Apparatus for the XUV-pump-XUV-probe studies and the temporal characterization on an isolated XUV burst (see SI). The XUV-broadband coherent continuum radiation is generated in Xe gas jet (GJ1) by means of an IPG device. The XUV radiation after the reflection on a Si plate passes through an aperture and a thin Sn filter selecting the desired spectral range. Subsequently is focused into the target gas jet (GJ2) by using a gold split spherical mirror XUV-autocorrelator (**b**) The spectrum (average of 300 pulses) of the radiation used in the experiment. In the interaction region, this radiation can support isolated pulses with duration as short as ≈ 420 asec and intensities exceeding $10^{14}$ W/cm$^2$. Focusing this radiation in a Xenon gas target the charge states of $Xe^+$ and $Xe^{2+}$ have been observed in the TOF spectrum (**c**). The green lines are the border lines of orange filled areas.

**Figure 2. Non-linear process used for the temporal characterization of isolated XUV pulses and the probing of ultrafast evolving atomic coherences.** Schematic diagram of the multi-XUV-photon ionization process induced in Xenon by the interaction with a sequence of the two mutually delayed intense broadband XUV pulses. The two non negligible competing channels leading to double ionization are: I) two photon direct double ionization (TPDDI) in part through the AIS manifold and II) two photon sequential double ionization (SDI). At the 1 fs or even sub-fs pulse duration level the direct process dominates.

**Figure 3. XUV-pump-XUV-probe measurement of one femtosecond scale electron dynamics** (**a**) Second order autocorrelation trace retrieved by recording the dependence of the $Xe^{2+}$ signal on the delay between the XUV pulses. The grey-dashed line corresponds to the raw data. To make the trend of the trace clearer the running

averages of the raw data are taken over nine delay points (green-yellow-filled area). Two domains are clearly distinguishable in the trace. The interval around zero delay (-3 *fs* to 6 *fs*) serves for the XUV pulse characterization and the trace at longer delays (6 fs to 92 fs) traces the evolution of the induced atomic coherences. The purple-line-shaded curve is the Gaussian fit on the raw data in the area, where the two XUV parts are temporally overlapped. (**b**) An expanded area of the trace in A. The Gaussian fit on the raw data yields to an $\tau_{XUV} = 1.5^{+0.2}_{-1.1}\ fs$. **(c)** FT of the raw data of trace A for delays > 6 fs reviling the frequency differences of the excited states. The frequency resolution, resulting from the maximum delay between the two mutually delayed pulses, is ~0.03 $fs^{-1}$. The assignment of the numbered peaks can be found in the SI.

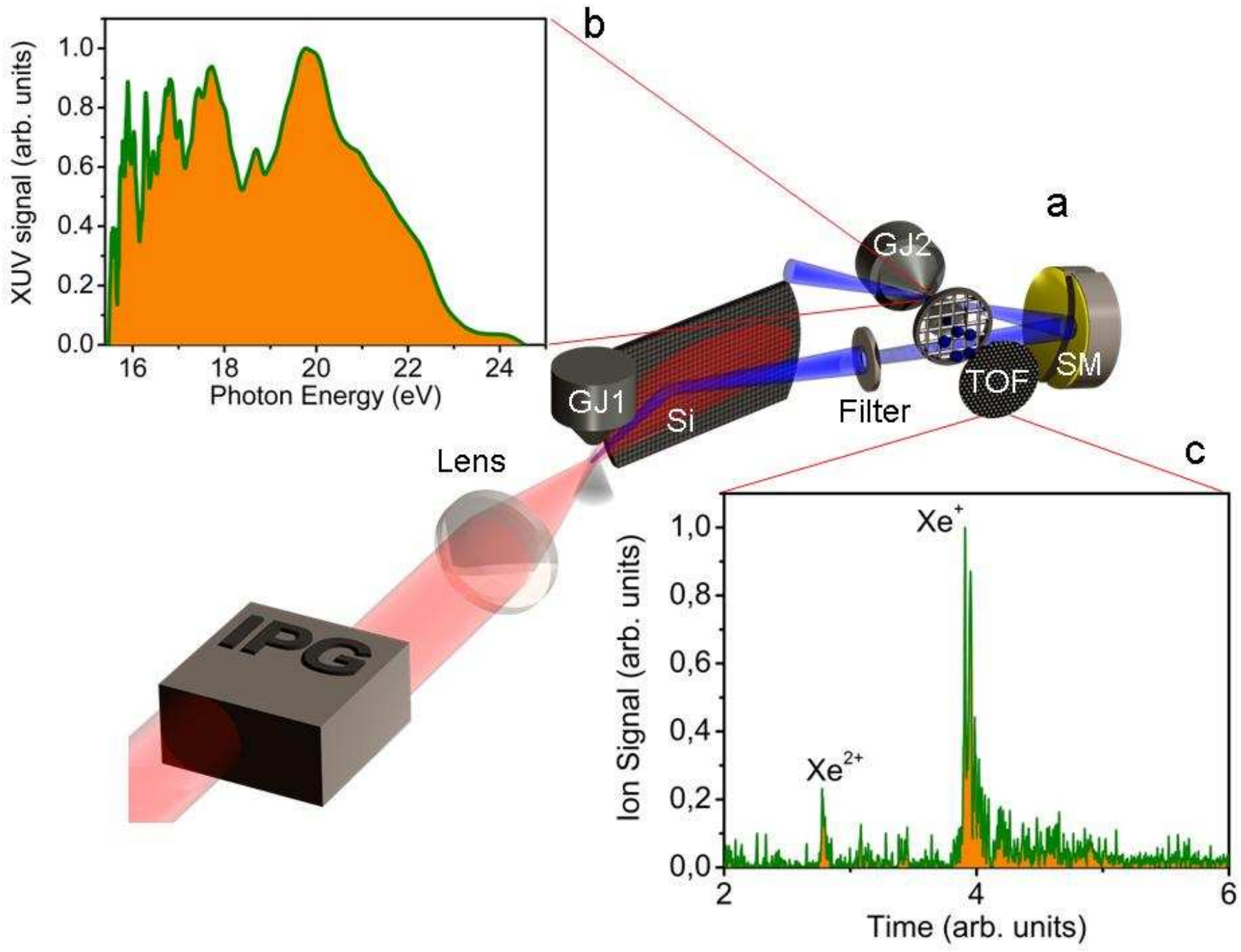

**Figure 1.**

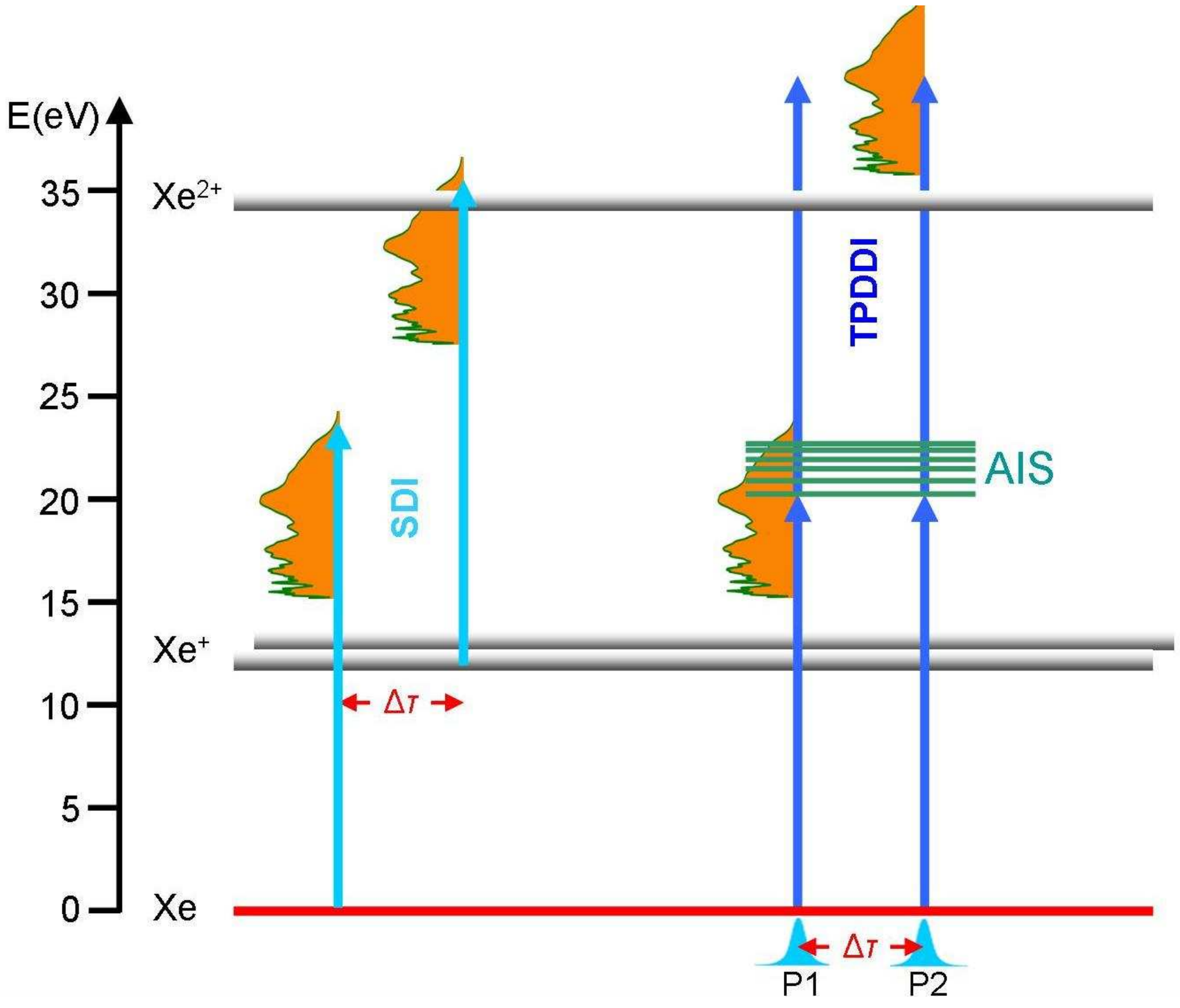

**Figure 2**

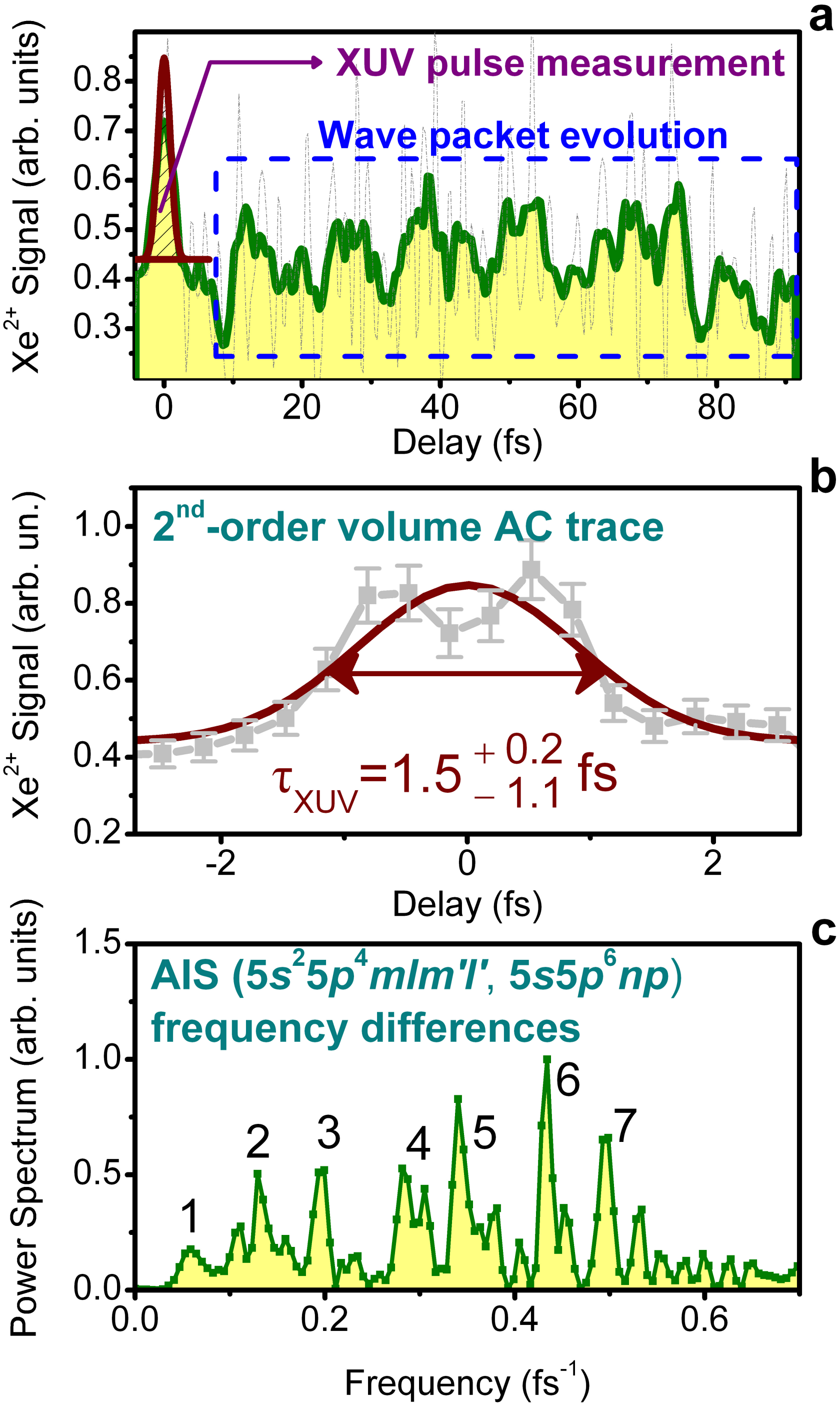

**Figure 3**

Supplementary Information

Supplementary information for:

# "Extreme-ultraviolet pump-probe studies of one femtosecond scale electron dynamics"

P. Tzallas[1*], E. Skantzakis[1,2], L. A. A. Nikolopoulos[3], G. D. Tsakiris[4] and D. Charalambidis[1,2]

[1]*Foundation for Research and Technology—Hellas, Institute of Electronic Structure and Laser, PO Box 1527, GR-711 10 Heraklion, Crete, Greece*

[2]*Department of Physics, University of Crete, PO Box 2208, GR71003 Heraklion, Crete, Greece*

[3]*School of Physical Science, Dublin City University, Glasnevin, Dublin 9, Ireland*

[4]*Max-Planck-Institut für Quantenoptik, Hans-Kopfermann-Str. 1, D-85748 Garching, Germany*

*Corresponding author e-mail address: ptzallas@iesl.forth.gr

**Experimental procedure and results**

The experiment was performed utilizing a 10 Hz repetition rate Ti:sapphire laser system delivering pulses up to 170 mJ/pulse energy at $\tau_L$=38 fs duration and wavelength at 800 nm (IR). The experimental setup which is shown in Fig. 1a of the manuscript is also described in detail elsewhere [1, 2]. The laser beam was passing through a Double Mach-Zender Interferometric Polarization Gating (DMZ-IPG) device (IPG in Fig. 1a of the manuscript). The technique is complementary to other polarization gating methods based on the wave-plate scheme [3-5], the two-color optical gating [6, 7] or the collective effects in XUV generation process [8, 9]. The DMZ-IPG output ellipticity modulated laser beam was focused with an $f$ =3 m lens into a pulsed gas jet (GJ1 in Fig. 1a of the manuscript), filled with Xe where the XUV broadband coherent continuum radiation was generated. After the jet a Si plate was placed at Brewster's angle of 75° of the fundamental, reflecting the harmonics [10] toward the detection area, while substantially reducing the IR field. The XUV radiation after reflection from the Si plate passes through a 3 mm diameter aperture and a 150 nm thick Sn filter in order to select the central part of the beam, and the required spectral region with a central wavelength of ≈ 60 nm, while eliminates the residual part of the IR beam. Subsequently the XUV pulse was split in two halves and focused into the target gas jet (GJ2 in Fig. 1a of the manuscript) by a split spherical gold mirror (SM in Fig. 1a of the manuscript) of 5 cm focal length. The reflectivity of gold is approximately constant along the bandwidth of the XUV radiation passing through the Sn filter. The autocorrelator delay was introduced by translating one of the two halves of the split mirror. The split mirror was mounted on a tilting stage and its movable half was mounted on a z, θ, φ translation-tilting feedback-controlled piezo-crystal translation stage. The minimum displacement step of the z-translation piezo-crystal unit is 1.5 nm. Great care has been taken of the proper spatiotemporal overlap of the two XUV beam halves at the focus of the split mirror. The spectrum of the XUV radiation was determined by measuring the energy-resolved, single-photon ionization, photoelectron spectra of Ar gas. The electron spectra were recorded using a μ-metal shielded time-off-flight (TOF) ion/electron spectrometer. Figure SI-1 shows typical XUV spectra recoded for the combinations where the GATE in the IPG device was ON and the Sn filter was IN the XUV beam line (red line), the GATE was OFF and the Sn filter was IN (blue line) and the GATE was ON and Sn filter OUT (black line). In the last case (Filter-OUT, GATE-ON) an XUV coherent continuum radiation reaching wavelengths as short as ≈ 25 nm has been recorded (black line). The spectral region from ≈ 80 nm to ≈ 50 nm was selected (red line) by placing the Sn filter in the XUV beam. The green line is the transmission curve of the Sn filter. The purple-doted line is the product of the transmission curve of the Sn filter times the spectrum recorded with the GATE is ON and the Sn filter OUT. The resulted transmitted XUV spectrum (dotted-purple line) is in reasonable agreement with the experimental one (red line).

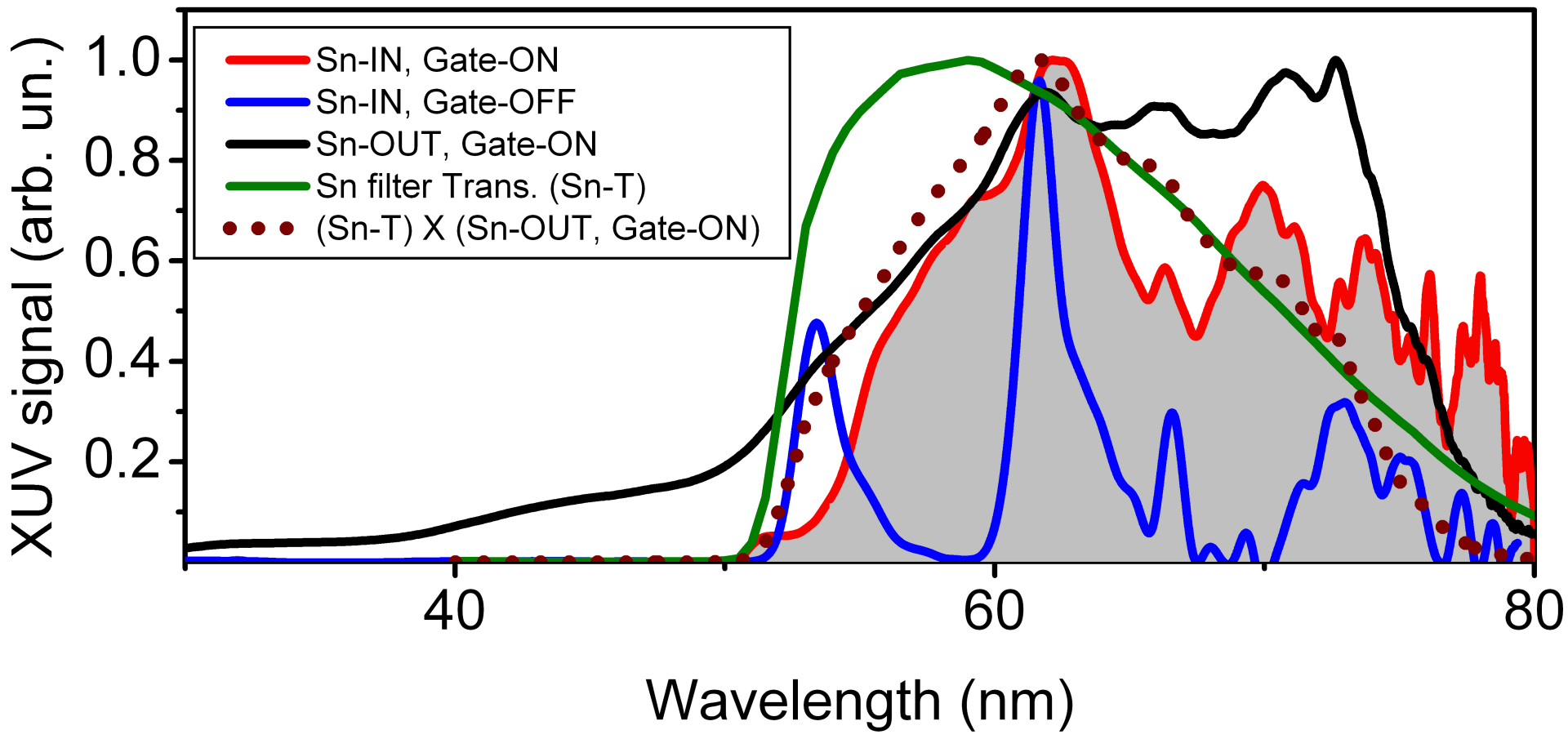

**Figure SI-1.** Red line: XUV spectrum recoded when the GATE in the IPG device is ON and the Sn filter is IN the XUV beam line (Sn-IN, Gate-ON). Blue line: XUV spectrum recoded when the GATE is OFF and the Sn filter is IN (Sn-IN, Gate-OFF). Black line: XUV spectrum recoded when the GATE is ON and Sn filter OUT (Sn-OUT, Gate-ON). Green line: Sn filter transmission curve (Sn-T). Purple-doted line: The spectrum has been calculated by multiplying the transmission of the Sn filter with the spectrum recorded with the GATE ON and the Sn filter OUT ((Sn-T)x(Sn-OUT, Gate-ON)).

The spectrum of the radiation used in the experiment is shown in Fig. 1b of the manuscript. It supports isolated pulses with duration as short as ≈ 420 asec. The energy of the reflected by the Si plate XUV radiation was measured to be ~100 nJ. The intensity in the interaction region is estimated to be between $3.2 \times 10^{13}$ and $2.5 \times 10^{14}$ W/cm$^2$. The high XUV intensity allowed the recording of the non-linear dependence of the $He^+$ on the XUV intensity (Figure SI-2). Figure SI-2a shows a measured time-of-flight mass spectrum. The peaks with m/z = 40, 18 and 2 are attributed to the single-XUV-photon ionization of Ar, $H_2O$ and $H_2$ respectively, while the peak with m/z = 4 is attributed to the two-XUV-photon ionization of He. The $H_2O$ and $H_2$ were background gases, while Ar gas has been introduced in the He gas line for the calibration of the XUV intensity scale. Figure SI-2b shows the dependence of the $He^+$, $H_2O^+$ and $H_2^+$ on the XUV intensity. The slope of 2 ± 0.2 for $He^+$ ascertains the two-XUV-photon ionization process for He, while the slopes of 0.7 ± 0.3 and 1 ± 0.3 for $H_2O^+$ and $H_2^+$, ascertain the linear single-XUV-photon ionization process for $H_2O$ and $H_2$.

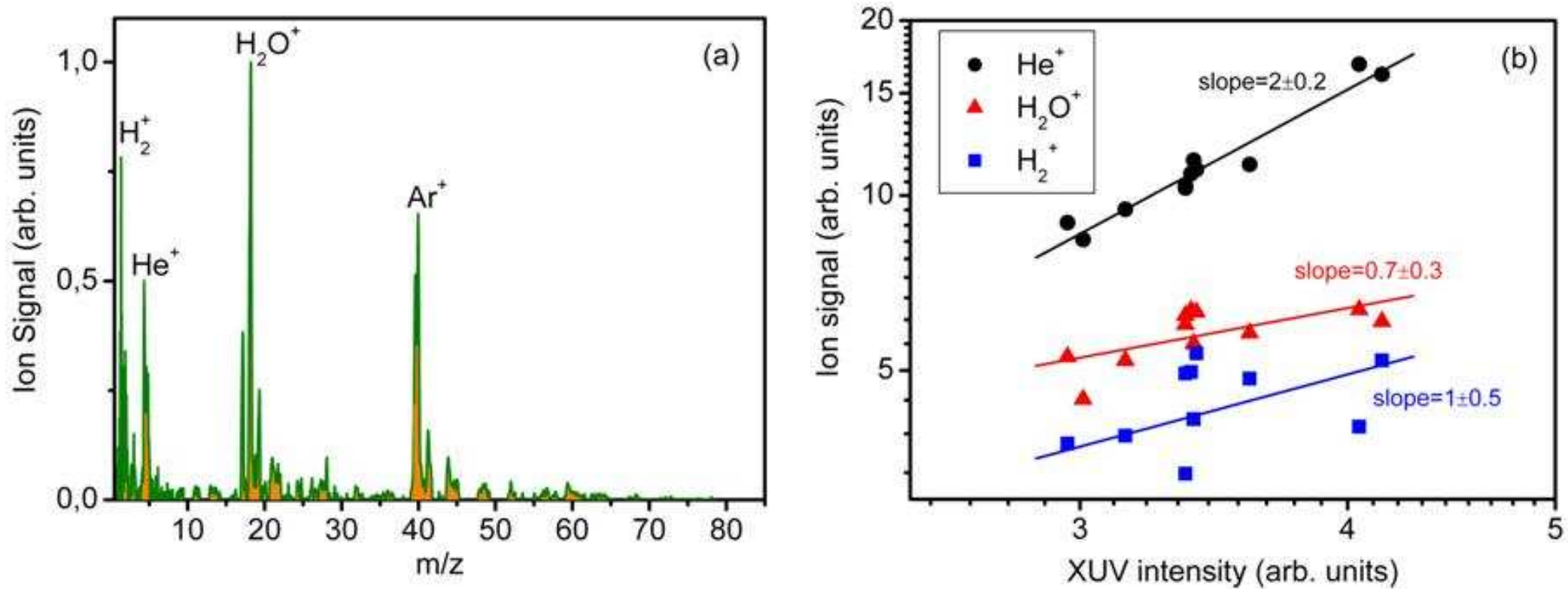

**Figure SI-2. (a)** Measured time-of-flight mass spectrum recorded ionizing with intense coherent continuum XUV radiation. In the spectrum the peaks with m/z = 40, 18 and 2 are attributed to the single-XUV-photon ionization of Ar, $H_2O$ and $H_2$ respectively, while the peak with m/z = 4 to the two-XUV-photon ionization of He. The $Ar^+$ yield being proportional to the XUV intensity serves calibrating the XUV intensity scale. **(b)** Dependence of the $He^+$, $H_2O^+$ and $H_2^+$ ion yields on the intensity of the XUV radiation. The slope of 2 ± 0.2 for $He^+$ verifies a two-XUV-photon He ionization, while the slopes of 0.7 ± 0.3 and 1 ± 0.3 for $H_2O^+$ and $H_2^+$ confirm single-XUV-photon ionization of $H_2O$ and $H_2$ respectively.

Further non-linear XUV-radiation-atom interactions are demonstrated in Xe gas and exploited in the investigation of induced ultra fast evolving atomic coherences and in second order autocorrelation based pulse metrology. In the TOF mass spectrum produced ionizing Xe with the intense XUV radiation (Fig. 1c of the manuscript) appear both $Xe^{+}$ and $Xe^{2+}$ ion peaks. Double ionization is only possible by at least two-XUV-photon absorption. Single photon absorption is partially resonant with the AIS manifolds and/or leads to single ionization leaving the ion in its ground or few excited states. Thus a coherent superposition of several AIS is excited. At the same time, since the XUV pulse duration is shorter than the time interval in which the spin-orbit occurs (pulse duration shorter than the precession period) the two fine structure levels of $Xe^{+}$ are coherently populated. Lower lying excited states of $Xe^{+}$ could also be coherently excited but the spectral amplitude of the exciting radiation is very low in this energy region. Those ultra-fast evolving atomic coherences are probed and the second order autocorrelation of the XUV pulse is obtained by recording the $Xe^{2+}$ signal as a function of the displacement between the two halves of the split mirror. For each delay step 100 data points were accumulated.

The produced radiation is intense enough to induce a two-photon ionization. Lowest-order perturbation theory predicts an ionization law, from a bound initial state, as $\sigma^{(N)} \cdot (I/\omega)^{N} \cdot t_N$ where $\sigma^{(N)}$ is the generalized cross section, I the intensity, ω the photon energy, $t_N$ the effective field duration and N the number of photons absorbed. In the present case, the lowest-order channel (in terms of number of photon absorptions) for the production of $Xe^{+3}$ is the direct 3-photon absorption from the Xe ground state with photons having at least energy around 21.5 eV and above. The XUV intensity dependence of $Xe^{2+}$ does not show any signature of saturation (fully depleted ground state). The lack of saturation is supported by the lack of higher charge states of Xe.

One may think that intensities close to $10^{14}$W/cm$^2$ would be sufficient for a three-photon ionization which would lead to the production of an observable signal of $Xe^{3+}$ ions. But from the fact that only part of the field spectrum is available for $Xe^{+3}$ production (above 21.5 eV) and that the corresponding 3-photon cross section is low we conclude that the intensity of the pulse in the experiment was not sufficiently enough for an observable $Xe^{+3}$ signal. This observation is also supported from the rather small measured $Xe^{+2}$ signal and thus makes clear why no higher charge states are present in the mass spectra. Indicating cross section values, dictated from the 2-and- 3-photon single-electron ionization cross sections, may be $10^{-51}$ cm$^4$ sec and $10^{-83}$ cm$^6$ sec$^2$ but without an estimate of their accuracy even in terms of order of magnitude.

**Assignment of the beating frequencies**

The assignment of the peaks of Fig. 3c is shown in Table 1. Note that the peaks correspond to frequency differences of excited states and not to excitation frequencies that are too high to be observed by the available temporal resolution. Each frequency peak is an overlap of few non-resolved spectral components. All frequency components can be resolved through a substantial increase of the total length of the temporal delay.The correlation between the $Xe^{2+}$ and $Xe^{+}$ temporal traces is fairly visible in their FT spectra. The FT spectrum of $Xe^{+}$ is shown in ref. 5 of the main manuscript. Within the frequency error of ± $30 \times 10^{-3}$ fs$^{-1}$ resulted by the temporal sampling window, the FT spectrum of $Xe^{2+}$ includes the frequency differences of states appearing in the FT spectrum of $Xe^{+}$ too. Additionally, the coherence induced through single photon ionization of Xe in the superposition of the two fine structure levels $P_{1/2}$ and $P_{3/2}$ of the ground state of the $Xe^{+}$, having an energy difference of 1.3eV (beat frequency 0.31 $fs^{-1}$) [Ref. 6 of main manuscript] probed through the sequential double ionization contributes to peak No. 4 of Fig. 3c. To the $Xe^{2+}$ trace may further contribute beating frequencies corresponding to the energy differences of the Xe ionic manifolds $5s5p^6$ and $5s^25p^46s$ and $5s^25p^45d$ coherently populated via single photon ionization of Xe. Those frequencies may contribute to the peaks No. 2, 4, 5 and 6 of fig. 3c. However, the XUV spectral amplitude is small to negligible in this energy region. In this sense peak No. 6 could not be firmly assigned. It does not appear in the $Xe^+$ FT spectrum of ref. 5 and thus should be attributed to coherently excited ionic states. Its frequency interval fits to few frequency differences between the $5s5p^6$ state and components of the $5s^25p^45d$ manifold, lying $\geq$ 25eV above the Xe ground state, an energy region, where the XUV spectral distribution has negligibly small amplitude.

**Table 1:** Numbers 1-10 in parenthesis stand for doubly excited states with energies in eV 20.664, 20.805, 21.03, 21.407, 21.721, 22.333, 22.457, 22.514, 22.56, 22.617, respectively. The energy values of the $5s[^2S_{1/2}]5p^66p$-$5s[^2S_{1/2}]\ 5p^612p$ inner-shell exited states are 20.952, 22.226, 22,704, 22.937, 23.068, 23.151, 23.206 eV. The Xe ionic state manifolds considered are the $5s5p^6$, $5s^25p^46s$ , $5s^25p^45d$ ones

*This peak includes also the frequency of the splitting of the two fine structure levels of the ground state of $Xe^+$.

| Peak No. | ($\nu$ ($fs^{-1}$) ± 30) x$10^{-3}$ | State pairs contributing to the peak<br>AIS manifolds | Ionic manifolds |
|---|---|---|---|
| 1 | 55 | ($5s5p^611p$, $5s5p^69p$)<br>($5s5p^66p$, 2)<br>($5s5p^68p$, 8-9)<br>(10,7)<br>(6, 8)<br>(2, 1) | |
| 2 | 129 | ($5s5p^611p$-$5s5p^610p$, $5s5p^6\ 8p$)<br>($5s5p^6\ 7p$, 10) | ($5s5p^6$, $5s^25p^4\ (^3P_2)6s$) |
| | | (5s5p$^6$ 8*p*, 6)<br>(5s5p$^6$ 9*p*, 8-9)<br>(4, 3)<br>(5s5p$^6$12*p*, 8*p*)<br>(5s5p$^6$ 7*p*,5)<br>(5s5p$^6$10*p*, 8-9) | (5s5p$^6$, 5s$^2$5p$^4$ ($^3$P$_2$)5d$_{3/2,5/2}$ [2])<br>(5s5p$^6$, 5s$^2$5p$^4$ ($^3$P$_2$)5d$_{7/2}$ [3])<br>(5s5p$^6$, 5s$^2$5p$^4$ ($^3$P$_1$)5d$_{1/2}$ [1]) |
| 3 | 200 | (5s5p$^6$ 9*p*, 5s5p$^6$ 7*p*)<br>(5s5p$^6$ 6*p*, 5)<br>(5s5p$^6$ 10*p*, 6)<br>(5s5p$^6$ 11*p*, 7)<br>(5s5p$^6$11*p*, 9-10)<br>(5s5p$^5$ 12*p*, 7-8)<br>(7, 5)<br>(4, 1)<br>(5s5p$^5$ 7*p*, 5s5p$^5$ 11*p*-10*p*)<br>(5s5p$^5$ 12*p*, 6)<br>(5, 2)<br>(4, 5-6)<br>(5, 10) | |
| 4 | 296 | (5s5p$^5$ 6*p*, 5s5p$^5$ 7*p*)<br>(5s5p$^5$ 6*p*, 6)<br>(5s5p$^5$ 8*p*, 4)<br>(5s5p$^5$ 10*p*, 5)<br>(6, 3) | (5s5p$^6$, 5s$^2$5p$^4$ ($^3$P$_2$)5d$_{7/2,9/2}$ [4])<br>(5s5p$^6$, 5s$^2$5p$^4$ ($^3$P$_0$)6s$_{1/2}$[0])<br>((5p$^5$)P$_{3/2}$,(5p$^5$)P$_{1/2}$) |
| 5 | 360 | (5s5p$^5$ 6*p*, 7-10)<br>(5s5p$^5$ 7*p*, 1)<br>(5s5p$^5$ 9*p*, 4)<br>(2, 6-7)<br>(3, 9-10) | (5s5p$^6$, 5s$^2$5p$^4$ ($^3$P$_1$)6s$_{3/2}$[1]) |
| 6 | 440 | | (5s5p$^6$, 5s$^2$5p$^4$ ($^3$P$_2$)5d$_{3/2}$ [1])<br>(5s5p$^6$, 5s$^2$5p$^4$ ($^3$P$_2$)5d$_{1/2}$ [0])<br>(5s5p$^6$, 5s$^2$5p$^4$ ($^3$P$_2$)5d$_{5/2}$ [3])<br>(5s5p$^6$, 5s$^2$5p$^4$ ($^3$P$_1$)6s$_{1/2}$[1]) |
| 7 | 500 | (5s5p$^5$ 6*p*, 5s5p$^5$ 10*p*-12*p*)<br>(5s5p$^5$ 9*p*, 1-2)<br>(5s5p$^5$ 10*p*, 2)<br>(5s5p$^5$ 11*p*, 3)<br>(5s5p$^5$ 12*p*, 3) | |

**The calculated temporal trace of the helium double ionization**

Helium is an up-scaled in energy (Figure SI-3a) and of much simpler structure system with a similar behavior at higher frequencies than Xe. These calculations are meant to show the general features of such an experiment and are not to provide a quantitative comparison. At the same time it provides the ground for pulse metrology at shorter times (larger bandwidths). The results are summarized in Figure SI-3b, which includes both high frequency components and cycle averaged traces.

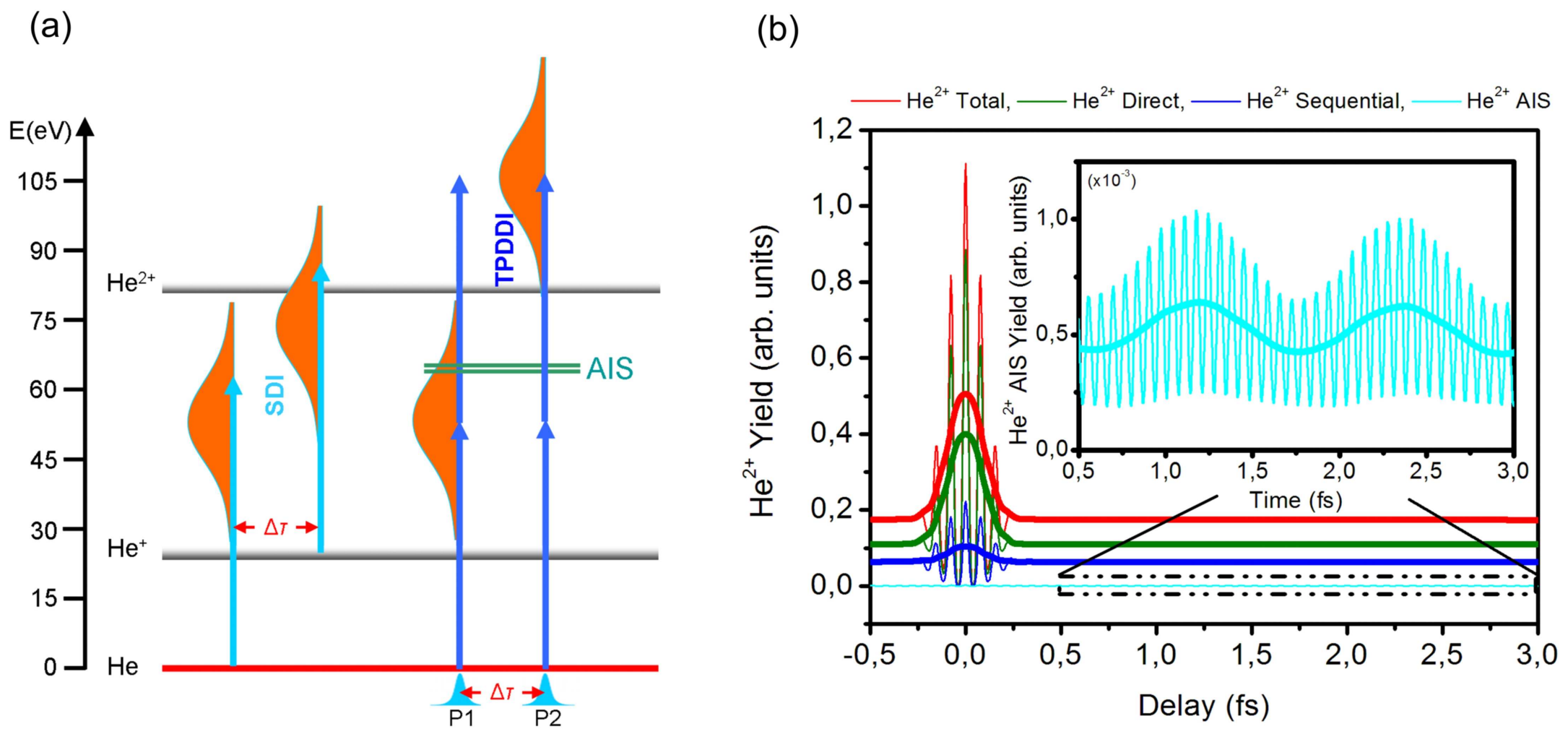

**Figure SI-3.** Calculated dependence of $He^{2+}$ yield on the time delay between two isolated attosecond pulses of 200 asec duration, 53 eV central photon frequency and $10^{14}$ W/cm$^2$ intensity . **(a)** Schematic diagram of the multi-XUV-photon ionization process induced in Helium by the interaction with a sequence of the two mutually delayed intense broadband XUV pulses. **(b)** The dependence of the total $He^{2+}$ yield on the delay is shown in red and is equivalent to the conventional 2$^{nd}$ order interferometric autocorrelation trace. The total $He^{2+}$ yield is the sum of the $He^{2+}$ yield resulting from the TPDDI of He (green line) in part through the AIS (2s2p, 2s3p) (light blue line) and the two-photon SDI of He (dark blue line). The thick lines are the corresponding cycle averaged traces. The energy difference between the 2s2p and 2s3p AIS is reflected to the slow modulation period shown in the inset.

The red line corresponds to the total $He^{2+}$ yield. For overlapping pulses, at small delays, the pronounced feature is mainly due to the direct double ionization channel (green line), i.e. the second order autocorrelation of the pulses, with a smaller contribution of the sequential double ionization (dark blue line). The trace of the sequential process is the square of the first order autocorrelation of the pulse and thus coincides in width with that of the direct process only for FTL pulses. For longer delays the high and low frequency beating originating from the excitation control and the evolution of the coherent superposition of the AIS is observable (light blue line). Note that for the used spectral distribution only two AIS are excited by the far high energy tail of the spectral distribution. A complete analogy with the Xenon case would demand for an even shorter pulse since appreciable population of the AIS requires a pulse of duration that is smaller than the inverse of the smaller energy difference between the various AIS ( $\tau_{XUV} < 1/\Delta E_{ij}$ ). By recalling that in Xe $\Delta E_{ij}$= 0.3 eV while in the present case $\Delta E_{12}$ ~3.5 eV and in conjunction with the smaller transition amplitudes in helium results to the observed much lower fringe amplitudes in He.

For the calculations we have employed a time-dependent perturbative approach of the appropriate order and the helium atomic structure calculated through an ab-initio configuration interaction (CI) method. For the calculation of the dependence of the $He^{2+}$ yield on the delay between the two isolated attosecond pulses the following formulation has been used. The photon energy of the field is sufficient to ionize helium from its ground state raising one of the electrons in the ($^1$P) continuum with kinetic energy $Q_1$ and simultaneously exciting both electrons onto He (2s2p), He(2s3p) autoionization states from which it can decay into the continuum $|c_1\rangle$, due to the electron-electron interaction. The energies of the 2s2p and 2s3p autoionizing states are 60.137 eV and 63.64 eV, respectively. Furthermore the system can absorb one more photon, reaching beyond the double ionization threshold, either through absorption from the $He^+$ ground state (sequential path) or from the exited virtual state (direct path) or from the autoionizing states $|a_1\rangle = |2s2p\rangle$ and $|a_2\rangle = |2s3p\rangle$. The Hamiltonian of the system is $H = h_0 + D(t)$ where $h_0$ is the field-free helium Hamiltonian and the free E/M field respectively. The operator $D(t) = [E(t) + E(t+t_d)]\hat{e}r$ represents the interaction between the atom and the field in the dipole approximation. In the latter expression, $t_d$ is the time delay between the two fields. For clarity, we have adopted a Fano formalism for the continuum and the autoionization states, which follows closely the formalism adopted in ref. 11. For the given field intensity and the frequencies (contained in the pulse) it be can easily shown that time-dependent perturbation theory is perfectly valid. The most general form of the time-dependent state of the system will be of the form:

$$|\psi(t)\rangle = C_g(t)|g\rangle + \sum_{i=1,2} C_{a_i}(t)|a_i\rangle + \sum_{i=1,2} \int d\varepsilon_{c_i} C_{c_i}(E_{c_i}, t)|c_i\rangle \quad (1)$$

where with $|g\rangle$ and $a_i$ being the He ground and the autoionization states of helium of the $^1$P symmetry, respectively. The single-ionization continuum of helium is represented as $c_1$ while the double ionization as $c_2$. Our task is to determine the dependence on the time delay of the state $\psi(t)$ through the amplitudes $C_g(t), C_a(t), C_{c_1}(t), C_{c_2}(t)$, which allow us to calculate the temporal trace of the doubly charged helium ion signal. The dynamics of the system is governed by the TDSE which reads $id\psi(t)/dt = [h_0 + D(t)]\psi(t)$. Inserting Eq. (1) into the latter TDSE and following a standard procedure for integrals involving over continua we obtain a system of first order integro-differential equations for the amplitudes of the state vector $\psi(t)$ (which in the present context are completely equivalent with the corresponding density matrix equations):

$$i\dot{C}_g(t,t_d) = [\bar{\varepsilon}_g - i(\gamma_1+\gamma_2)/2]C_g(t,t_d) + \sum_{i=1,2} D_{ga_i}C_{a_i}(t,t_d)$$

$$i\dot{C}_{a_i}(t,t_d) = [\bar{\varepsilon}_{a_i} - i\gamma_{a_i}/2]C_{a_i}(t,t_d) + D_{ga_i}C_g(t,t_d) + \sum_{i'\neq i} \Omega_{ii'}C_{a_i}(t,t_d)$$

$$i\dot{C}_{c_1}(t,t_d) = [\bar{\varepsilon}_{c_1} - i\gamma_{12}/2]C_{c_1}(t,t_d) + \sum_{i=1,2} V_{c_1a_i}C_{a_i}(t,t_d) + D^{(1)}_{c_1g}C_g(t,t_d)$$

$$i\dot{C}_{c_2}(t,t_d) = \bar{\varepsilon}_{c_2}C_{c_2}(t,t_d) + D_{c_21}C_{c_1}(t,t_d) + \sum_{i=1,2} D_{c_2a_i}C_{a_i}(t,t_d) + D^{(2)}_{c_2g}C_g(t,t_d)$$

In the above equations, $\gamma_1(t)$, $\gamma_2(t)$ represent the instantaneous ionization width of the ground state to the singly-ionized and doubly-ionized helium. $\gamma_{a_i}(t)$ $i$ = 1, 2 are the instantaneous widths of the autoionization states $a_i$ to (a) singly-ionized helium and to (b) doubly-ionized helium through the absorption of one-extra photon from the field. Finally, $\gamma_{12}(t)$ represents the ionization width of the singly-charged ion to the doubly-charged ion. The definitions for the one-photon dipole matrix elements $D^{(1)}_{gc_1}(t)$, $D_{ga_1}(t)$, $i$ = 1, 2, $D_{gc_1}$ as well as the electron-electron interaction matrix elements $V_{a_ic_1}$, $i$ = 1, 2 are given in [11]. All the single-photon ionization widths are proportional to the field intensity while the dipole transition amplitudes to the electric field. The two-photon ionization width ($\gamma_2$) is proportional to the squared of the direct two-photon transition amplitude $D^{(2)}_{gc_2}(t)$ has a $I^2$ dependance on field intensity. An important element of the observed signal is its dependence on the field-dependent amplitudes $\Omega_{ii'}(t)$ which represent Raman couplings between the autoionization states through all single-photon accessible states. Finally, in the energies above of the relevant states $(\bar{\varepsilon}_g, \bar{\varepsilon}_{a_i}, \bar{\varepsilon}_{c_1})$ we have taken into account all the associated dynamic Stark shifts. All the above transition amplitudes and ionization widths of helium are ab-initio calculated through a multiconfiguration correlation interaction (CI) approach, developed over the years the details of which can be found in [12].

Integration of the above amplitudes equations (for fixed time delay between the pulses each time) will give ionization probabilities to the various states as a function of the time delay $t_d$.

The $He^{2+}$ yield has been calculated in the presence of a two temporally delayed FTL broadband XUV pulses of 200 asec duration and intensity $10^{14}$ W/cm$^2$. The electric field $E(t)$ is taken to have maximum at photon frequency $\omega$ = 53 eV, with a Hanning-like temporal shape with bandwidth at FWHM of about 21 eV.

**References**

1) Skantzakis, E. *et al.*, *Opt. Lett.* **34**, 1732 (2009).
2) Charalambidis, D. *et al.*, *New J. Phys.* 10, 025018 (2008).
3) Tcherbakoff, O. *et al.*, *Phys. Rev.* **A 68**, 043804 (2003).
4) Oron, D. *et al.*, *Phys. Rev. A* **72**, 063816 (2006).
5) Altucci, C. *et al.*, *Opt. Lett.* **33**, 2943 (2008).
6) Mashiko, H. *et al.*, *Phys. Rev. Lett.* **100**, 103906 (2008).
7) Takahashi, E. et al., *Phys. Rev. Lett.* 104, 233901 (2010)
8) Strelkov, V. V. Mevel, E. Constant, E. *New J. Phys.* 10, 083040 (2008).
9) Ferrari, F. *et al.*, *Nature Phot.* **4,** 875 (2010).
10) Takahashi, E. J. *et al.*, *Opt. Lett.* **29**, 507 (2004).
11) Nikolopoulos, L. A. A. *et al.*, *Eur. Phys. J. D* **20**, 297 (2002).
12) Nikolopoulos, L. A. A. Lambropoulos, P. *J. Phys. B* **34**, 545 (2001).